\documentclass[preprint,aps,prl,groupedaddress,showpacs]{revtex4-1}

\usepackage{graphicx}
\usepackage{bm}
\usepackage{amsmath}
\usepackage[bottom]{footmisc}

\begin{document}

\title{Nonlinear Transfer of Intense Few Cycle Terahertz Pulse Through Opaque n-doped Si}

\author{O.\,V.~Chefonov}

\author{A.\,V.~Ovchinnikov}

\author{M.\,B.~Agranat}

\author{V.\,E.~Fortov}
\affiliation{Joint Institute for High Temperatures of the Russian Academy of Sciences (JIHT RAS), Izhorskaya 13 Bldg. 2, Moscow 125412, Russian Federation}

\author{E.\,S.~Efimenko}

\author{A.\,N.~Stepanov}
\affiliation{Institute of Applied Physics of the Russian Academy of Sciences (IAP RAS), Ul'yanov Street 46, Nizhny Novgorod, 603950, Russian Federation}

\author{A.\,B.~Savel'ev}
\email[Corresponding author: ]{abst@physics.msu.ru}
\affiliation{Lomonosov Moscow State University, Leninskie Gory 1, bld.62, Moscow, 119991, Russian Federation}

\date{\today}

\begin{abstract}
Intense few cycle terahertz pulses exhibit complex non-linear behavior under interaction with heavily n-doped Si. Fast increase in the transmission of a 700~fs pulse (central frequency $\sim1.5$~THz) through the Si sample (low field transmission of $\sim0.02$~\%) saturates  at $\sim8$~\% for the external field of $5$~MV/cm and then drops twofold at $\sim20$~MV/cm. An electro-optical sampling measurements revealed formation of a single cycle terahertz pulse at this field  due to formation of a thin ionized layer by the first intense oscillation of the terahertz field. 
\end{abstract}




\maketitle

\section{Introduction}

An ultra-strong terahertz (THz) electric-field has become available for experiments recently and this opened feasibility of nonlinear terahertz optics. Record-breaking results have been obtained using synchrotron radiation sources (pulse energy $W_{\text{THz}}>100 \mu$J, electric-field strength $E_{\text{THz}}$ up to 20~MV/cm in the frequency range $>10$~THz) \cite{ref1} or using optical rectification in nonlinear organic crystals ($W_{\text{THz}}\sim$900~$\mu$J,  $E_{\text{THz}}\sim$50~MV/cm in the frequency range $\sim1$~THz)~\cite{ref2, ref3, ref4, ref5}. These values are enough for tackling nonlinearity of a medium in optical range, and it also depends on the wavelength  (quiver electron energy, etc.) that is $\sim$100 times higher in the THz domain than in the optical one.

Self-induced transparency of a medium under action of an electromagnetic radiation is one of the most important nonlinear effects in the optical range~~\cite{ref6, ref7}, including that observed in semiconductors~\cite{ref8}. In the terahertz range this effect (named bleaching) was previously observed in semiconductors irradiated with a THz field of relatively low strength~\cite{ref9, ref10, ref11}, and increase in transmission was of a few percent only. Recently, the giant enhancement of transmission from 0.03 to 2.5\% in n-doped Si was observed at a maximum THz field  $E_{\text{THz}}\sim$3.1~MV/cm~\cite{ref12}. This effect was qualitatively interpreted within the Drude model because of increasing collision rate under action of the THz field.

In this work we observed that transmission by energy of the 700~fs THz pulse through n-doped Si sample amounts to as high as $\sim8$\% with field strength up to 5--7~MV/cm and then gradually decreases nearly twofold at higher electric fields of 10--20~MV/cm. Analysis of the transmitted pulse waveform demonstrated its strong distortions and generation of higher-frequency spectral components at 7--10~MV/cm, and finally a 300-fs single-cycle pulse was observed at maximum field. Simulations showed that with $E_{\text{THz}}\sim$5--7~MV/cm saturation of the electron-phonon collision rate occurs, and THz transmission saturates. Generation of a single-cycle pulse at even higher fields was associated with formation of a thin ionized layer at the front sample's surface during the first intense oscillation of the THz field cutting out subsequent field oscillations and decreasing total transmission.

\section{Experimental layout and methods}

Intense THz pulses were generated by optical rectification of femtosecond laser pulses delivered by the Cr:forsterite laser system (1240~nm, 100~fs, 10~mJ, 10~Hz) in the organic nonlinear crystal DSTMS (4-N,N-dimethylamino-4'-N'-methyl-stilbazolium 2,4,6-trimethylbenzenesulfonate) of 8~mm in diameter~\cite{ref5}.  A telescope 6:1 consisted of two off-axis parabolic mirrors with focal length of 25.4 and 152.4~mm was used to expand a THz beam up to $\sim28$~mm FWHM. The THz beam was focused to a 200-$\mu$m spot (FWe$^{-1}$H) using an off-axis parabolic mirror with focal length and diameter equal to 50.8~mm. Experimental layout is described in detail in~\cite{ref12}. 

Terahertz pulse energy $W_{\text{THz}}$ was measured by a calibrated Golay cell (GC-1D Tydex). This energy was controlled by changing the pump laser energy with an attenuator consisted of a half-wave plate and Glan-Thompson prism. Measurements of the transverse mode of the THz beam were made both moving a 200-$\mu$m aperture across the beam and producing damage pattern of a thin metal film ~\cite{ref13}. The THz spot was nearly circular at the focal plane, and its line-out was Gaussian-like with a radius of 110~$\mu$m (ÍWHM). By the contrast, an elliptical spot with 1:2 axial ration was observed at displacement $z$ out of the focal plane by $\sim 1.3$~mm (the beam waist $\sim0.55$~mm). Here the spot has flat top central part and broad wings with amplitude up to 10\%. 

\begin{figure}[!ht]
\includegraphics{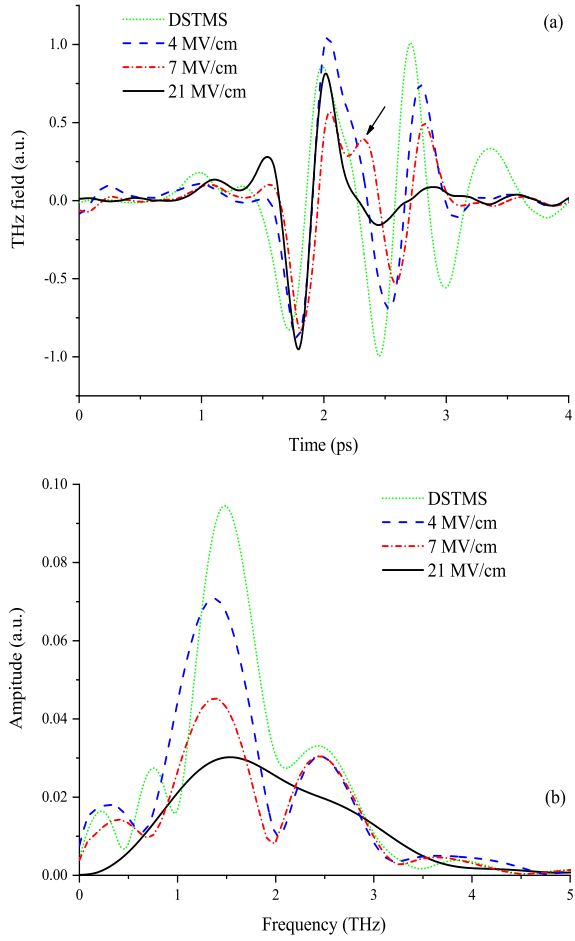}	
\caption{\label{fig1} (a) Waveforms and (b) Fourier-transformed spectra of THz pulses without (DSTMS) and with the sample at $E_{\text{THz}} \sim 4.3, 7.1, 21$~MV/cm. Arrow in Fig.~\ref{fig1}~(a) points at the specific bend in the curve for $E_{\text{THz}}\sim7.1$~MV/cm.}
\end{figure}

The time-domain electric field waveform was obtained using electro-optical sampling (EOS) in a 200-$\mu$m thick gallium phosphide crystal (GaP) with femtosecond laser radiation at 1240~nm as a probe. It should be noted that optical axes of the THz and the probe beams were collinear, while their spots were 200 and 50~$\mu$m in diameter at the crystal's surface. Waveform and amplitude spectral density of the initial THz pulse with duration of 700~fs (FWHM) and central frequency of 1.5~THz are shown in Fig.~\ref{fig1} (a), (b). Note, that the waveform remains unchanged at different pump laser energies.

Experiments were carried with a n-doped Si wafer 245~$\mu$m thick, with carrier concentration of $9\times10^{16}$cm$^{-3}$ and mobility of 800~cm$^2$/V (from the Hall effect measurements) that corresponds to collision rate of $1.1\times10^{13}$~s$^{-1}$. To measure sample transmission $\eta$ as a function of the strength $E_{\text{THz}}$ two approaches were utilized: (i) the  THz pulse energy was adjusted by changing pump laser energy at a fixed sample position or (ii) open aperture $z$-scan technique was employed at a fixed $W_{\text{THz}}$ value.

\section{Experimental results}

\begin{figure}[!ht]
\includegraphics{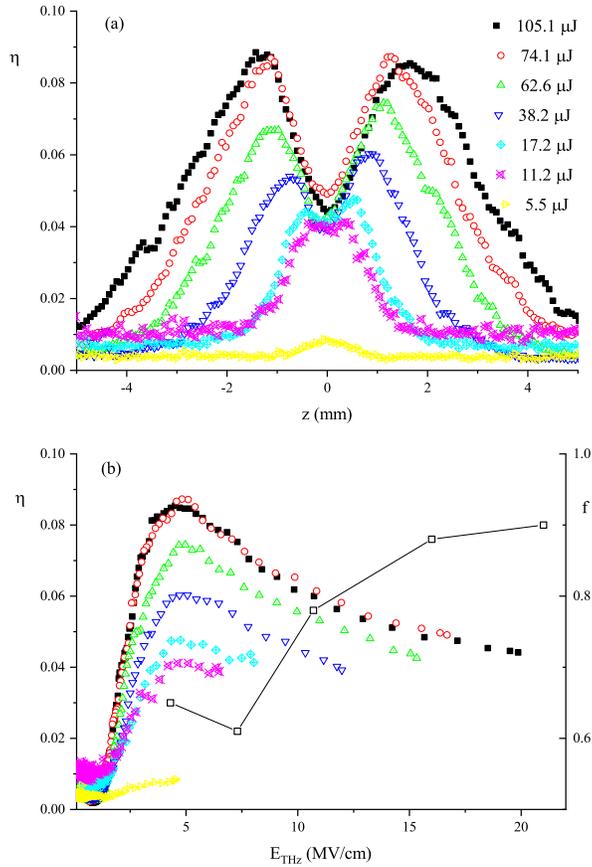}
\caption{\label{fig2} Experimental dependencies of the transmission $\eta$ on (a) displacement $z$ from the focal point and (b)  strength $E_{\text{THz}}$ at different energies $W_{\text{THz}}$. Squares in Fig.~\ref{fig2} (b) display a fraction $f$ of the THz pulse energy after the sample within the first field oscillations.}
\end{figure}

Experimental dependences of the transmission $\eta$ from $z$-scan measurements are shown in Fig.~\ref{fig2} (a). The bleaching is clear at  \textquotedblleft low\textquotedblright energy
$W_{\text{THz}}\sim11$~$\mu$J with maximum transmission $\eta_{\text{max}}\sim4$\% reached at the exact focal point $z$=0. Contrary,  the transmission reached maximal value of 8.5\% out of the focal point ($z=\pm1.2$~mm) at higher energies $W_{\text{THz}}$.

Dependencies of the THz transmission $\eta$ on the field strength $E_{\text{THz}}$ are plotted in Fig.~\ref{fig2} (b). Curves were calculated using $z$-scan data from Fig.~\ref{fig2} (a) follow algorithm described in~\cite{ref12} on the assumption of ideal transverse Gaussian intensity distribution. Note, that the observed changes in the THz spot shape along $z$ introduce ambiguity in determination of the strength $E_{\text{THz}}$ from the displacement $z$. One can see that $\eta_{\text{max}}$ reaches $\sim8.5$\% at $E_{\text{THz}}\approx 5$~MV/cm, then it decreases nearly twofold at electric fields higher than 10--15~MV/cm. Note, that increase of the $W_{\text{THz}}$ from 74 to 105~$\mu$J did not result in noticeable changes of the transmission curve (see Fig.~\ref{fig2} (b)). 

\begin{figure}[!ht]
\includegraphics{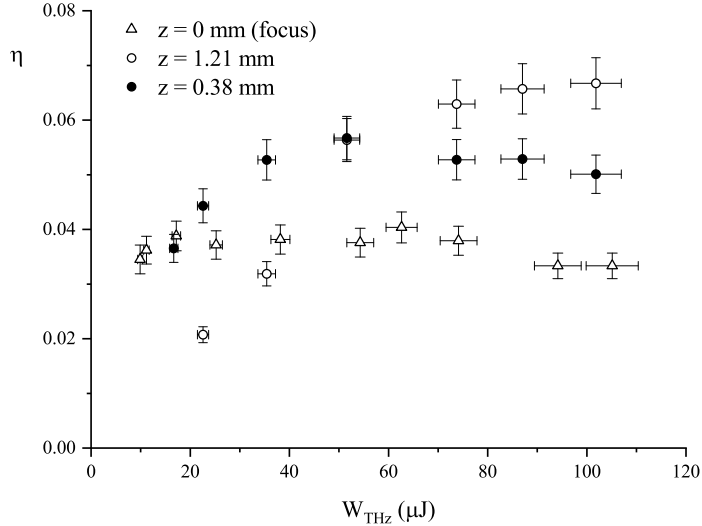}
\caption{\label{fig3} Experimental dependence of the transmission $\eta$ on the energy $W_{\text{THz}}$ at different displacements $z$.
}
\end{figure}

Additionally, dependencies $\eta(W_{\text{THz}})$ were measured at three sample's positions (at focus of the mirror, at maximum transmittance, $z$=1.2~mm, and at the middle point $z$= 0.38~mm) to validate the unusual effect of maximal transmittance out of the focal position (Fig.~\ref{fig3}).  The data obtained in this way completely supported $z$-scan data: $\eta_{\text{max}}$ was reached at $\sim1.2$~mm offset from the focal plane .

Waveforms and Fourier-transformed spectra of transmitted THz pulses at various  $E_{\text{THz}}$ are shown in Fig.~\ref{fig1}. Measurements were carried out with 
$W_{\text{THz}}\approx 100$~$\mu$J, while the field strength was changed by changing the displacement $z$. The waveform and the spectrum of the transmitted pulse at $E_{\text{max}} = 4$~MV/cm are very similar to the incident ones. Substantial changes to the THz pulse waveform and spectrum were observed with field increase up to 7--10~MV/cm. The spectral amplitude around 1.5 THz begins to fall while the spectral amplitude around 2.4~THz remains almost unaltered that results in a waveform curve bend (marked with an arrow in Fig.~\ref{fig1} (a)). Finally at 21~MV/cm, the output pulse is a single-cycle one with a sharp rising edge, spectrum broadened up to 4~THz and central frequency shifted from 1.5 to 1.8~THz.

\section{Discussion}

The data obtained could be understood from competition of processes of bleaching and ionization under action of an intense THz field. For carrier concentration around $10^{17}$~cm$^{-3}$ bleaching occurs due to increase in the collision rate $\nu$  of electrons with acoustic phonons \cite{ref14}.  There are no experimental and numerical data on collision and ionization rates in Si for a (quasi) dc electric field above 1~MV/cm because of the breakdown of a Si  \cite{ref15}. Hence short ($<$~1~ps) and intense THz pulses provide a unique possibility to investigate different materials at quasi dc fields unavailable other way.

Quantum-mechanical calculations of the collision rate  $\nu(\varepsilon)$ as a function of electron energy $\varepsilon$  were presented in \cite{ref16, ref17}. This rate rapidly increases in the energy range 0.5--1.5~eV, then remaining practically constant ($\sim1.2-1.5\times10^{14}$~s$^{-1}$) up to energies of 5~eV. Hence a relation between the instant field strength $E$ and electron energy $\varepsilon$ should be established.

The collision rate $\nu$ significantly exceeds the central circular frequency $\omega$ of the THz pulse at $E>0.5$~MV/cm and consequently at each instant an electron velocity equals to the drift velocity 
$v_d = e E/ m \nu(\varepsilon)$, where $e$, $m$ --- electron charge and effective mass and  $\varepsilon = 0.5mv^{2}_{d}$. 

An intense THz field could induce generation of non-equilibrium carriers~\cite{ref18} as a result of the impact ionization by free carriers oscillating in an external field~\cite{ref19, ref20, ref21} or the tunnel ionization (Zener effect)~\cite{ref22, ref23}. Silicon is non-direct-gap semiconductor and tunnel ionization probability is negligible. Impact ionization is possible if electron energy substantially exceeds the bandgap (1.12~eV for Si at 300~K). One can find several approximating formulae for the impact ionization rate in silicon resulting in quite similar curves \cite{ref20, ref21}.

\begin{figure}[!ht]
\includegraphics{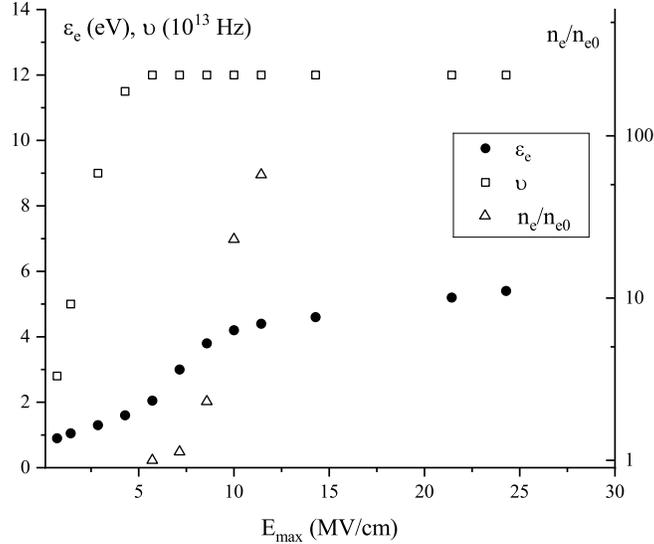}
\caption{\label{fig4} Calculated dependencies of the maximum electron energy $\varepsilon_{\text{e}}$ (squares), electron-phonon collision rate $\nu$ (circles) and electron concentration $n_{\text{e}}$ (triangles) on the  $E_{\text{THz}}$,  $n_{\text{e0}} = 9\times10^{16}$~cm$^{-3}$.}
\end{figure}

Calculated dependencies of energy $\varepsilon$, rate $\nu$ and concentration of additional electrons $n_{\text{e}}$ on the maximum field strength $E_{\text{THz}}$ (inside a sample) are shown in Fig.~\ref{fig4} . The collision rate increases at $E_{\text{THz}} < 5$~MV/cm and then reaches a constant value determined by our approximation of $\nu(\varepsilon)$ dependence. It results in an increase of carriers' energy up to several eV and at $E_{\text{THz}} \approx 7$~MV/cm this energy becomes high enough to induce rapid increase of an electron density.

An estimate of THz transmission $\eta$ for a plane wave using Fresnel formulae with Drude dielectric permittivity and $\nu \sim 1.2-1.5\times10^{14}$~s$^{-1}$ gives $\eta\sim$~8--9\% well corresponding to the experimentally observed value. Therefore, the experimentally observed saturation of bleaching with increase in $E_{\text{THz}}$ to 5~MV/cm could be attributed to saturation of the collision rate $\nu$.

We calculated dependence of the fraction $f$ of the THz pulse energy contained within the first THz field oscillation (0.8--2.5~ps) regarding to the $W_{\text{THz}}$  on the strength $E_{\text{THz}}$ (calculated using experimental data plotted in Fig.~\ref{fig1}) is shown in Fig.~\ref{fig2} (b). This fraction slightly exceeds 0.5 for the maximum bleaching, and it increases up to 0.9 at higher fields. This change causes a nearly two-fold decrease in transmission from 8\% to 4\% observed in our experiments. 

To make deeper insight into the spectral-temporal changes of the THz pulse, preliminary numerical simulations of nonlinear propagation of an intense THz pulse through a silicon plate were performed. The Finite Difference Time Domain method was used to solve numerically the 1D Maxwell's equations for the electromagnetic field together with equations for plasma currents and carrier concentration. Dependencies of both the collision and impact ionization rates in Si on the $E_{\text{THz}}$ were taken into account when calculating the current. 

\begin{figure}[!ht]
\includegraphics{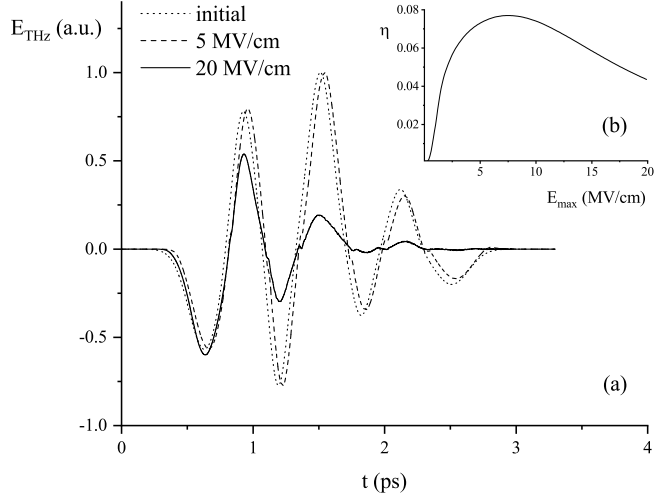}
\caption{\label{fig5} (a) Waveforms of THz pulses before (dotted line) and after propagation through the Si sample at the $E_{\text{THz}} = 5$ (dashed line) and 20 (solid line) MV/cm. (b) Inset shows  transmission dependence on the $E_{\text{THz}}$.}
\end{figure}

Propagation of an intense THz pulse through the sample causes significant changes in the pulse waveform and spectra with prominent suppression of all oscillations except for the first one (Fig.~\ref{fig5}), that qualitatively corresponds to the observed experimental data. A thin ionized layer is formed at the input surface of the sample during the first oscillation of the THz field with $E_{\text{THz}}>7$~MV/cm. An electron density inside this layer drops from $10^{19}$~cm$^{-3}$ to $10^{18}$~cm$^{-3}$ within a few microns  at the highest filed strength of 21~MV/cm. In spectral domain it causes an increase of the plasma frequency and predominant dumping of low-frequency spectral components. In temporal domain the oscillations of the THz field except for the first one are suppressed due to an increase in reflectance (from 0.3 to 0.4) and absorption. All these processes resulted in a single-cycle THz pulse with a sharp rising edge and smooth spectrum in the range of 0.5--4 THz (see Fig.~\ref{fig1}). 

The inset to the Fig.~\ref{fig5} (b) also shows calculated dependence of the transmission $\eta$ on the $E_{\text{THz}}$ for the Gaussian beam. This curve qualitatively agreed with experimental ones obtained using $z$-scan technique (Fig.~\ref{fig2} (b)): the transmission reaches maximum at 5--10~MV/cm due to collision rate increase and saturation and then decreases slowly when ionization becomes important. Contrary, experimental data obtained with the sample fixed at the certain position $z$ (Fig.~\ref{fig3})  demonstrates saturation of the transmission after reaching the maximum value. We attribute this difference to the discussed above changes in the transverse THz field distribution along the pulse propagation. Note, that  the experimentally measured transmission is averaged over the whole beam cross-section. Simple estimations showed, that averaged transmission obtained for the Gaussian beam should be smaller than for the flat-top beam of the same strength $E_{\text{THz}}$ by a factor of 1.3--1.5.

\section{Conclusion}

In summary, we studied the interaction of an intense THz pulse with the n-doped Si sample under the conditions inaccessible to quasi-stationary electric fields. With increasing maximum electric-field strength above 1~MV/cm a complex nonlinear dynamics of transmission of the 245-$\mu$m thick n-doped Si with carrier concentration of $9\times10^{16}$~cm$^{-3}$  was observed. An increase of electron-phonon collision rate (due to an increase of ''instantaneous'' electron energy) results in the bleaching at electric fields below 5~MV/cm. The maximum transmission by energy was found out $\sim8$\% that is more than two orders of magnitude higher than with a low-intensity THz pulse. Saturation of the transmission is attributed to saturation of the collision rate at electron energies above 1.5~eV. The transmission gradually decreases at higher fields above 7~MV/cm that is related to the onset of ionization by electron impact. Ionization occurs in a thin surface layer during the first intense oscillation of the THz field. This results in efficient reflection and attenuation of subsequent field oscillations leading to formation of a single-cycle THz pulse with sharp rising edge and broad spectrum. This pulse propagates deeper into the sample under the bleaching conditions. The observed effect strongly depends on the spatial energy distribution over the THz beam cross section and, in our experiments, the maximum transmittance was observed out of the focal plane.

Mutual orientation of a crystallographic axes of a sample and polarization of THz radiation seems of crucial interest for a further research, as well as angle of incidence dependence. The n-doped Si can be used as an effective saturable absorber in the THz domain. Here peak transmission and its contrast under the action of low- and high-intensity THz fields can be controlled by thickness and carrier concentration of a sample. Finally, such a n-Si sample could be used to control THz pulse shape and spectrum, and even generate a single-cycle THz pulses with a sharp rising edge and a broad spectrum.

\begin{acknowledgments}
Authors wish to thank T.\,Ozaki for numerous helpful discussions. The work was supported by Russian Science Foundation no. 17-19-01261. Experiments were performed using the unique scientific facility ''Terawatt Femtosecond Laser Complex'' in the Center ''Femtosecond Laser Complex'' of the Joint Institute for High Temperatures of the Russian Academy of Sciences.
\end{acknowledgments}


\providecommand{\noopsort}[1]{}\providecommand{\singleletter}[1]{#1}%

\end{document}